\documentclass[submission,copyright,creativecommons]{eptcs}
\providecommand{\event}{GandALF 2025} % Name of the event you are submitting to
\usepackage{iftex}
\ifpdf
  \usepackage{underscore}         % Only needed if you use pdflatex.
  \usepackage[T1]{fontenc}        % Recommended with pdflatex
\else
  \usepackage{breakurl}           % Not needed if you use pdflatex only.
\fi

\usepackage{listings}
\usepackage{subfigure}
\usepackage{xspace}
\usepackage{graphicx}
\usepackage[utf8]{inputenc}
\usepackage{multirow}
\usepackage{enumitem}
\usepackage{rotating}
\usepackage{float}

\usepackage{algpseudocode}
\usepackage{graphicx}
\usepackage{amsmath}
\usepackage{mathtools}
\usepackage{algorithm}
\usepackage{syntax}
\usepackage{amssymb}
\usepackage{array}
\usepackage{mathrsfs}
\usepackage{listings}
\usepackage{xcolor}
\usepackage{euscript}

\newcommand{\powerset}{\EuScript{P}}

\lstdefinelanguage{ISPL}{
  morekeywords={Agent, Vars, Actions, Protocol, Evolution, Evaluation, Groups, Formulae, end},
  commentstyle=\color{gray},
  basicstyle=\ttfamily\scriptsize,
  morecomment=[l]{//},
}

\lstdefinelanguage{AgentSpeak}{
  morekeywords={+!,+,-,!,?,.,if,then,else},
  morecomment=[l]{//},
  morestring=[b]",
  alsoletter={.,!},
  commentstyle=\color{gray}\ttfamily,
  stringstyle=\color{orange},
  basicstyle=\ttfamily\small,
  literate={->}{{$\rightarrow$}}1
           {<=}{{$\Leftarrow$}}1
           {<-}{{$\leftarrow$}}1
           {=>}{{$\Rightarrow$}}1
           {not}{{$\neg$}}1,
}

	\usepackage{stackrel}
\newcommand{\lnotation}[4]{
	\def\third:{#3} 
	\def\possiblyone:{} 
	\def\possiblytwo:{~}
	\def\possiblythree:{ }
	\def\divide{\;#1\hspace*{-0pt}( #2\; \mid: \; #4 \, )}
	\def\nodivide{\;#1\hspace*{-0pt}( #2\;\mid\; #3\;:\;#4 \, )}
	\ifx\third\possiblyone\divide
		\else\ifx\third\possiblytwo\divide
		\else \ifx\third\possiblythree\divide
		\else \nodivide\fi\fi\fi}

\newcommand{\biglnotation}[4]{
	\def\third:{#3} 
	\def\possiblyone:{} 
	\def\possiblytwo:{~}
	\def\possiblythree:{ }
	\def\divide{\;#1\hspace*{-0pt}\big( #2\; \mid: \; #4 \, \big)}
	\def\nodivide{\;#1\hspace*{-0pt}\big( #2\;\mid\; #3\;:\;#4 \, \big)}
	\ifx\third\possiblyone\divide
		\else\ifx\third\possiblytwo\divide
		\else \ifx\third\possiblythree\divide
		\else \nodivide\fi\fi\fi}

\newcommand{\bigglnotation}[4]{
	\def\third:{#3} 
	\def\possiblyone:{} 
	\def\possiblytwo:{~}
	\def\possiblythree:{ }
	\def\divide{\;#1\hspace*{-0pt}\bigg( #2\; \mid: \; #4 \, \bigg)}
	\def\nodivide{\;#1\hspace*{-0pt}\bigg( #2\;\mid\; #3\;:\;#4 \, \bigg)}
	\ifx\third\possiblyone\divide
		\else\ifx\third\possiblytwo\divide
		\else \ifx\third\possiblythree\divide
		\else \nodivide\fi\fi\fi}

\newcommand{\grieslnotation}[4]{
	\def\third:{#3} 
	\def\possiblyone:{} 
	\def\possiblytwo:{~}
	\def\possiblythree:{ }
	\def\divide{(#1 #2\; \mid : \; #4 \, )}
	\def\nodivide{(#1 #2\;\mid\; #3\;:\;#4 \, )}
	\ifx\third\possiblyone\divide
		\else\ifx\third\possiblytwo\divide
		\else \ifx\third\possiblythree\divide
		\else \nodivide\fi\fi\fi}
	\usepackage{color}
	\usepackage{etoolbox}
	\usepackage{soul}
\definecolor{darkred}{rgb}{0.75,0.0,0.0}
\definecolor{darkgreen}{rgb}{0.0,0.6,0.0}
\definecolor{darkblue}{rgb}{0.0,0.0,0.6}
\definecolor{darkcyan}{rgb}{0.0,0.6,0.6}
\definecolor{darkmagenta}{rgb}{0.6,0.0,0.6}
\definecolor{darkamber}{rgb}{1.0,0.5,0.0}
\definecolor{darkyellow}{rgb}{0.6,0.6,0.0}

\definecolor{lightred}{rgb}{1.0,0.9,0.9}
\definecolor{lightgreen}{rgb}{0.9,1.0,0.9}
\definecolor{lightblue}{rgb}{0.9,0.9,1.0}
\definecolor{lightcyan}{rgb}{0.8,1.0,1.0}
\definecolor{lightmagenta}{rgb}{1.0,0.8,1.0}
\definecolor{lightamber}{rgb}{1.0,0.8,0.0}
\definecolor{lightyellow}{rgb}{1.0,1.0,0.8}

\definecolor{webgreen}{rgb}{0,0.5,0}
\definecolor{webbrown}{rgb}{0.6,0,0}

\definecolor{grey}{rgb}{0.65,0.65,0.65}
\definecolor{purple}{rgb}{0.4,0,0.75}

\definecolor{burgundy}{rgb}{0.5, 0.0, 0.13}         % For states
\definecolor{darkcyan}{rgb}{0.0,0.6,0.6}            % For sync messages
\definecolor{darkpastelgreen}{rgb}{0.01, 0.75, 0.24}% For guarded transitions

\newcommand{\mynote}[2]{
	\ifstrequal{#1}{0}{\textcolor{darkamber}{#2}}{}%
  	\ifstrequal{#1}{1}{\textcolor{darkmagenta}{#2}}{}%
  	\ifstrequal{#1}{2}{\textcolor{darkcyan}{#2}}{}%
  	\ifstrequal{#1}{3}{\textcolor{darkgreen}{#2}}{}%
  	\ifstrequal{#1}{5}{\textcolor{darkblue}{#2}}{}%
  	\ifstrequal{#1}{8}{\textcolor{burgundy}{#2}}{}
}
\newcommand{\todiscuss}[2]{
	\ifstrequal{#1}{0}{\textcolor{darkamber}{\textit{\textbf{TO DISCUSS}: #2}}}{}%
  	\ifstrequal{#1}{1}{\textcolor{darkmagenta}{\textit{\textbf{TO DISCUSS}: #2}}}{}%
  	\ifstrequal{#1}{2}{\textcolor{darkcyan}{\textit{\textbf{TO DISCUSS}: #2}}}{}%
  	\ifstrequal{#1}{3}{\textcolor{darkgreen}{\textit{\textbf{TO DISCUSS}: #2}}}{}
}
\newcommand{\todo}[2]{
	\ifstrequal{#1}{0}{\textcolor{darkamber}{\textbf{\underline{TO DO}}: #2}}{}%
  	\ifstrequal{#1}{1}{\textcolor{darkmagenta}{\textbf{\underline{TO DO}}: #2}}{}%
  	\ifstrequal{#1}{2}{\textcolor{darkcyan}{\textbf{\underline{TO DO}}: #2}}{}%
  	\ifstrequal{#1}{3}{\textcolor{darkgreen}{\textbf{\underline{TO DO}}: #2}}{}
}
\newcommand{\toaddress}[2]{
	\ifstrequal{#1}{0}{\noindent\textcolor{darkamber}{$\bigstar$~\textbf{#2}}\\}{}%
  	\ifstrequal{#1}{1}{\noindent\textcolor{darkmagenta}{$\bigstar$~\textbf{#2}}\\}{}%
  	\ifstrequal{#1}{2}{\noindent\textcolor{darkcyan}{$\bigstar$~\textbf{#2}}\\}{}%
  	\ifstrequal{#1}{3}{\noindent\textcolor{darkgreen}{$\bigstar$~\textbf{#2}}\\}{}
}
\newcommand{\goldseeker}{\textsl{Goldseeker}\@\xspace}

\newtheorem{definition}{Definition}

\newlist{tabitemize}{itemize}{1}
\setlist[tabitemize]{label=\textbullet, 
                     leftmargin=*,
                     nosep, 
                     before=\begin{minipage}[t]{\hsize}\raggedright, 
                     after=\end{minipage}}
\title{Generating Plans for Belief-Desire-Intention (BDI) Agents Using  Alternating-Time Temporal Logic (ATL)}
\author{
Dylan L\'{e}veill\'{e} 
\institute{Department of Systems and Computer Engineering\\
Carleton University, Ottawa, ON, Canada}
\email{dylan.leveille@carleton.ca}
}
\def\titlerunning{Generating Plans for BDI Agents Using ATL}
\def\authorrunning{D. L\'{e}veill\'{e}}
\begin{document}
\maketitle

\begin{abstract}
    Belief-Desire-Intention (BDI) is a framework for modelling agents based on their beliefs, desires, and intentions. Plans are a central component of BDI agents, and define sequences of actions that an agent must undertake to achieve a certain goal. Existing approaches to plan generation often require significant manual effort, and are mainly focused on single-agent systems.  As a result, in this work, we have developed a tool that automatically generates BDI plans  using Alternating-Time Temporal Logic (ATL). By using ATL, the plans generated accommodate for possible competition or cooperation between the agents in the system.  We demonstrate the effectiveness of the tool by generating plans for an illustrative game that requires agent collaboration to achieve a shared goal. We show that the generated plans allow the agents to successfully attain this goal.
\end{abstract}

% Paper Body
\section{Introduction}
\label{sec:introduction}
% Begin Section
% Computer systems are becoming increasingly intelligent and autonomous. 
While there exist many modern forms of Artificial Intelligence (AI), \textit{agents}  are a traditional form of AI that remain popular despite the growing trends of Machine Learning and Large Language Models. 
% There exists many definitions for agents~\cite{Ahmad2008}. 
An agent can be defined as ``anything that can perceive its environment through sensors, and act upon that environment through actuators based on its agent program''~\cite{russell2016artificial}. 
% Aligning with this definition, 
A popular framework for modeling agents is known as Belief-Desire-Intention (BDI)~\cite{Bratman1987-BRAIPA}. In this framework, agents are seen as having beliefs (i.e., knowledge) and desires (i.e., goals), which they use to justify their intention (i.e., actions)~\cite{Georgeff1995}. Planning theory is a fundamental concept in BDI. \textit{Plans} are hierarchical structures that represent possible future intentions for the agent~\cite{Bratman1987-BRAIPA}.  Put simply, intentions are the plans an agent adopts. An agent commits to a plan until is is completed. The generation of quality plans influences the agent's rationality. 
% The BDI framework has gained significant traction in the agent research community, and many works have been published on this topic~\cite{davoust2020architecture,braubach2004goal,sanchez2019designing}.

As BDI is only a conceptual framework, previous work has proposed concrete BDI-based programming languages that can be used to model real-world agents. One such language is AgentSpeak~\cite{rao1996agentspeak}. In AgentSpeak, the intentions of agents are modeled as linear plans, each prescribing a series of actions to be taken by the agent. 
% For a plan to be considered by the agent, the beliefs and desires of the agent must align with the pre-conditions of the plan itself. In other words, each plan is tied to a number of beliefs and desires that must be held by the agent for the plan to be considered. 
% Plans may also involve a change of beliefs and desires. 
While AgentSpeak is a powerful language for implementing BDI agents, plans must be generated manually. This activity is tedious and highly error-prone. In fact, in environments possessing many possible states, generating plans manually to achieve complex agent behaviour may not be feasible. A single mistake may cause undesirable behaviour, and can be difficult to debug~\cite{davies2024event}. 

To address this problem, we propose using Alternating-Time Temporal Logic (ATL) to generate AgentSpeak plans automatically for BDI agents. To achieve this, we have developed a tool that generates plans based on the actions an agent must take to satisfy a given ATL formula (representing the agent's goal). As the agent may be uncertain of any number of variables in its environment, plans are generated from all possible combinations of variable values. This results in plans  that account for any range of uncertainty from the agent.  By specifying goals using ATL formulas, the generated plans also consider potential competition or cooperation with other agents in the system. This work is novel, as the generation of BDI plans using ATL in multi-agent systems has not been previously explored.

% Academics are continuously exploring new methods to improve BDI agents. One such methods is with the use of formal methods. \textit{Formal methods} are mathematically-based techniques  for describing systems and  verifying the properties of those systems~\cite{Wing1990}. Formal methods is especially useful  Formal methods can take many forms, including first order logic. 

The rest of this paper is organized as follows. Section~\ref{sec:background} presents the necessary background information to understand the remainder of this work. Section~\ref{sec:relatedwork} discusses previous works on the topic of BDI plan generation. Section~\ref{sec:ATLBDI} motivates how ATL can be integrated into BDI agents. Section~\ref{sec:ATLantis} presents a tool to automatically generate AgentSpeak plans using ATL. Section~\ref{sec:uncertainty} demonstrates how the plans generated can be used practically at runtime. Section~\ref{sec:discussion} discusses the benefits of the plans generated, and possible limitations.  Lastly, Section~\ref{sec:conclusion} concludes and briefly discusses future work.
% End Section

\section{Background}
\label{sec:background}
% Begin Section
% In this section, we provide the necessary background information required to understand the rest of this work.

% \subsection{Belief-Desire-Intention}
% The Belief-Desire-Intention (BDI) framework was first proposed by Michael Bratman for modeling agent planning and decision-making~\cite{Bratman1987-BRAIPA}. In Bratman’s work, agents make decisions based on three key elements: \textit{beliefs}, \textit{desires}, and \textit{intentions}. These are defined as follows~\cite{Bratman1987-BRAIPA}:

% \begin{itemize}
%     \item Beliefs: A set of values believed to the true by an agent.
%     \item Desires: A set of objectives for which the agent exhibits ``pro-attitudes'' (i.e., are wanted, desirable by the agent).  
%     \item Intentions: Describes a commitment by the agent, which may involve performing certain actions, and the revision of certain beliefs, desires, and intentions of the agent.
% \end{itemize}

% Planning theory is a fundamental concept in Bratman's work. \textit{Plans} are hierarchical structures that represent possible future intentions for the agent~\cite{Bratman1987-BRAIPA}.  Specifically, intentions are plans that are realized by the agent.  The generation of quality plans influences the agent's rationality. 

\subsection{AgentSpeak}

\textit{AgentSpeak} is an abstract programming language developed by Rao~\cite{rao1996agentspeak} to model BDI agents. AgentSpeak provides the formal syntax required to define plans for BDI agents, and the operational semantics for the reasoning and execution of these plans. As it is not relevant for this work, we omit the formal syntax and semantic definitions for AgentSpeak. Instead, we present the general form of an AgentSpeak plan, which can be seen in Figure~\ref{fig:AgentSpeakForm}.

\begin{figure}[ht!]
\centering
\begin{lstlisting}[language=AgentSpeak, mathescape=true]
+!goal:
	guard$_\texttt{{1}}$ & guard$_\texttt{{2}}$ & ... & guard$_\texttt{{N}}$
	<-
	action$_\texttt{{1}}$; action$_\texttt{{2}}$; ... ; action$_\texttt{{M}}$.
\end{lstlisting}
\caption{The general form of an AgentSpeak plan}
\label{fig:AgentSpeakForm}
\end{figure}

% We formally define the syntax of AgentSpeak in the following definitions. Definition...\mynote{0}{finish rest of syntax definitions below: beliefs, goals, triggering events, actions, plans.} TELL THEM PLANS ARE LINEAR HERE.

% \begin{definition}[AgentSpeak Belief Syntax -- \cite{rao1996agentspeak}]
% \label{def:ASBeliefs}
% Let $b$ be a predicate, and $t_{1},...,t_{n}$ be terms where $n \geqq 1$. Then $b(t_{1},\dots ,t_{n} )$ is
% a belief atom. Let $c$ be a predicate, and $s_{1},...,s_{m}$ be terms where $m \geqq 1$. Then, $b(t_{1},\dots ,t_{n} ) \land s_{1},...,s_{m}$, and $\neg b(t_{1},\dots ,t_{n} )$ are beliefs.
% \end{definition}

As shown in the figure, a plan is linked to an agent’s goal and will only be considered when the agent holds that goal. To adopt the plan,  the plan's guard conditions must be satisfied. These typically represent the agent’s beliefs. Once adopted, the actions are executed in sequence. If multiple plans can be adopted, AgentSpeak will select one for execution. A well-known concrete implementation of AgentSpeak is the \textit{Jason} language. Jason is an AgentSpeak interpreter developed for Java which conforms to the syntax and semantics of AgentSpeak~\cite{bordini2007programming}. While AgentSpeak does not provide a mechanism for updating beliefs, Jason allows agents to update their individual beliefs based on perceptions from the environment. 
% \mynote{0}{Note: My assumption is that this is enough Jason needed for Background as my approach is much more related to AgentSpeak and BDI in general with respect to ATL, than Jason. }

\subsection{Alternating-Time Temporal Logic}
Alternating-Time Temporal Logic (ATL) is a branch of temporal logic that models agents, their actions, and the consequence of their actions on the environment~\cite{Alur1998}. It is primarily used for strategic reasoning. ATL can be interpreted as a multi-agent extension to Linear Temporal Logic (LTL) and Computation Tree Logic (CTL). Additionally, while LTL and CTL are for closed systems, ATL is intended for open systems (i.e., those in which the evolution of the environment depends on the actions of agents)~\cite{Alur1998}. 

ATL formulas are verified on Concurrent Game Models (CGM) or Concurrent Game Models with Incomplete Information (CGMII), which are defined in Definition~\ref{def:atlCGM} and Definition~\ref{def:atlCGMII} respectively.   The syntax of ATL is similar to that of LTL and CTL. Its syntax is defined in Definition~\ref{def:atlSyntax}. The semantics are slightly different to that of LTL and CTL, and are defined in Definition~\ref{def:atlSemantics}. 

\begin{definition}[Concurrent Game Model (CGM) -- \cite{Goranko2016}]
\label{def:atlCGM}
A Concurrent Game Model (CGM) $\mathcal{M}$ is defined by the following tuple:

\[
\mathcal{M} = (\Sigma, S, \Pi, Act, \textit{af},\textit{tf},\textit{vf})
\]

The elements in $\mathcal{M}$ are defined as:
\begin{itemize}
    \item $\Sigma$: The set of all agents in the system.
    \item $S$: The set of all possible environment states in the system.
    \item $\Pi$: The set of all propositions in the system.
    \item $Act$: The set of all agent actions.
    \item $\textit{af}$: An action function $\textit{af}: \Sigma  \times S \xrightarrow{} ( \powerset(Act) \setminus \emptyset )$, which assigns each agent at each state a set of possible actions.
    \item $\textit{tf}$: A transition function \textit{tf} which assigns to each state $s \in S$ and $\overrightarrow{\alpha}$ at $s$ (where $\overrightarrow{\alpha}$ is a tuple of actions $\overrightarrow{\alpha}=(a_{1},a_{2},...,a_{k} )$ with $a_{i} \in \textit{af}\:(i,s)$ and $i\in \Sigma$) a unique resulting state $\textit{tf}\:(s,\overrightarrow{\alpha}) \in S$.
    \item $\textit{vf}$: A valuation function $\textit{vf}: \Pi  \xrightarrow{}  \powerset(S) $, which assigns to each proposition the set of states for which it is true.
\end{itemize}

\end{definition}

\begin{figure}[ht!]
    \centering
    \centerline{\includegraphics[width=0.25\textwidth]{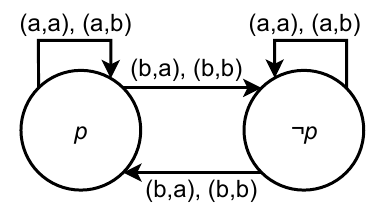}}
    \caption{An example CGM}
    \label{fig:cgmExample}
\end{figure}

An example CGM can be seen in Figure~\ref{fig:cgmExample}. In this example, states are drawn as nodes. Each state has a proposition, $p$, that is either \textit{true} or \textit{false}. Transitions are drawn using arrows between nodes. Note that there are two agents in the system, as each transition is activated by the independent actions of both. Let the first agent be \textit{Ag1}, and the second agent be \textit{Ag2}. For each transition, the action on the left of the tuple is the action performed by \textit{Ag1}, and the action on the right of the tuple is the action performed by  \textit{Ag2}. For example, $(a,b)$ implies \textit{Ag1} performed action $a$, while \textit{Ag2} performed action $b$. In this example, the transition from the state where $p$ is \textit{true} to the state where it is \textit{false} is activated if \textit{Ag1} and \textit{Ag2} perform actions $b$ and $a$ respectively, or actions $b$ and $b$ respectively. 

\begin{definition}[Concurrent Game Model with Incomplete Information (CGMII) -- \cite{Vester2013}]
\label{def:atlCGMII}
A Concurrent Game Model with Incomplete Information (CGMII) is an extension to CGM that accounts for possible uncertainty from each individual agent. In a CGMII, an agent can have a lack of visibility into its current state, making some states \textit{indistinguishable}. In these indistinguishable states, it may be difficult for an agent to determine the action that should be taken. A CGMII $\mathcal{M'}$ is defined by the following tuple:

\[
\mathcal{M'} = (\Sigma, S, \Pi, Act, \textit{af},\textit{tf},\textit{vf}, \{\sim  i\}_{i\in \Sigma})
\]

Where,
\begin{itemize}
    \item $(\Sigma, S, \Pi, Act, \textit{af},\textit{tf},\textit{vf})$: Is a CGM. 
    \item $\sim  i$: Is an equivalence relation for every agent $i \in \Sigma$ such that $\sim  i \subseteq S \times S$. It evaluates to true for two states that are indistinguishable for agent $i \in \Sigma$, and false otherwise. 
    \item For $s1, s2 \in S$, and $i \in \Sigma$, if $s1 \sim i \; \; s2$, then $\textit{af}\:(i,s1)=\textit{af}\:(i,s2)$.
\end{itemize}

\end{definition}

\begin{figure}[ht!]
    \centering
    \centerline{\includegraphics[width=0.25\textwidth]{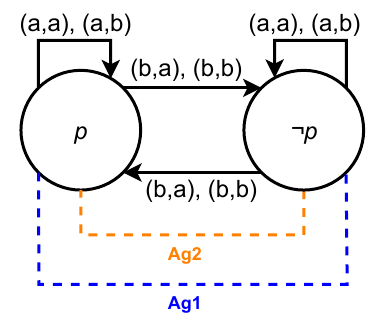}}
    \caption{An example CGMII}
    \label{fig:cgmiiExample}
\end{figure}

An example CGMII can be seen in Figure~\ref{fig:cgmiiExample}. Notice that this example is identical to that of Figure~\ref{fig:cgmExample}, except for the inclusion of dashed lines which represent the indistinguishability of both agents. Specifically, in this example, we say that both states are indistinguishable for agents \textit{Ag1} and \textit{Ag2} (i.e., both states appear identical to both agents). As the states are indistinguishable, this implies the agents have no visibility into the value of $p$. This lack of visibility into the value of $p$ makes it difficult for agents to determine the actions that would influence its value.

\begin{definition}[ATL Syntax -- \cite{VanDrimmelen2003}]
\label{def:atlSyntax}
Let $\Sigma$ be the set of all agents in a system, and $\Pi$ be the set of all propositions in the system. Additionally, let $A \subseteq \Sigma$ and $p\in\Pi$. An ATL formula $\varphi$ is defined as:

\[
\varphi ::= p \mid \neg \varphi \mid \varphi_1 \vee \varphi_2 \mid \langle\langle A \rangle\rangle X \varphi \mid \langle\langle A \rangle\rangle F \varphi \mid  \langle\langle A \rangle\rangle G \varphi \mid \langle\langle A \rangle\rangle \varphi_1 U  \varphi_2
\]

\end{definition}

\begin{definition}[ATL Semantics -- \cite{Goranko2016}]
\label{def:atlSemantics}
Unlike CTL and LTL, ATL formulas are used to determine whether a coalition of agents can adopt a strategy (i.e., a series of actions) to ensure a given condition holds. As a result, we notice that the temporal operators,  $X$,$F$, $G$, $U$, are accompanied by a coalition of agents. We define these coalition-augmented temporal operators as follows:
\begin{itemize}
    \item $\langle\langle A \rangle\rangle X \varphi$: used to verify whether there exists an action for coalition $A$  to enable $\varphi$ to be true in the immediate next state.
    \item $\langle\langle A \rangle\rangle G \varphi$: used to verify whether there exists a strategy for coalition $A$ to guarantee $\varphi$ to be true in the current state, and all future states. 
    \item $\langle\langle A \rangle\rangle F \varphi$: used to verify whether there exists a strategy for coalition $A$ to eventually allow $\varphi$ to be true in some future state.
    \item $\langle\langle A \rangle\rangle  \varphi_1 U \varphi_2$: used to verify whether there exists a strategy for coalition $A$ to ensure  $\varphi_1$ is true in the current state, and all subsequent states, until $\varphi_2$ becomes true. 
\end{itemize}

It is important to note that any agents specified in a coalition for a formula are assumed to be in \textit{cooperation} (i.e., agents in a coalition can work together to achieve a certain goal). Similarly, every other agent not specified in a coalition for a formula are assumed to be in \textit{competition} with the agents in the coalition (i.e., agents outside the coalition have no interest in assisting the agents within the coalition to achieve a certain goal). Lastly, note that strategies can be finite or infinite sequences of actions.

\end{definition}

As an example, suppose we wish to verify the ATL formula $\langle\langle Ag1 \rangle \rangle X(\neg p)$ on the CGM model given in Figure~\ref{fig:cgmExample}, with the initial state being the state where $p$ is \textit{true}. Informally, this formula is asking whether there exists an action for \textit{Ag1}, regardless of the action taken by \textit{Ag2}, to achieve $\neg p$. Given the initial state, there does exist a strategy: action $b$ will allow $\neg p$ to be \textit{true}, regardless of the action taken by \textit{Ag2}. Suppose we wish to verify the same formula on the CGMII model in Figure~\ref{fig:cgmiiExample}, with the initial state being the state where $p$ is true. In the state where $p$ is \textit{true}, the strategy is to perform action $b$. In the state where $\neg p$ is \textit{true}, the strategy is to perform action $a$. As \textit{Ag1} cannot distinguish between the two states, either action is considered a valid strategy, and the formula still holds. A formula holds in a CGMII if it is always attainable despite state indistinguishability, regardless of whether the required actions differ between the strategies.

Also note that a strategy is either \textit{uniform} or \textit{non-uniform}~\cite{Lomuscio2017}. For example, when verifying the ATL formula $\langle\langle Ag1 \rangle \rangle F(\neg p)$ on the CGM model given in Figure~\ref{fig:cgmExample}, with the initial state being the state where $p$ is \textit{true}, one strategy that satisfies this formula is to perform action $b$. However, many strategies exist, such as $a,b$, and $a,a,b$. As a result, each of these strategies are non-uniform. However, suppose we wish to verify the formula $\langle\langle Ag1 \rangle \rangle G(p)$ with the same CGM model and initial state. Only one strategy exists in this case; performing action $a$ indefinitely. As only one strategy exists, the strategy is uniform.
% End Section

\section{Related Work}
\label{sec:relatedwork}
% Begin Section
Plan generation for BDI agents has been extensively explored in the literature. One widely used approach is the creation of contingency plans~\cite{Meuleau2012}. These are tree-shaped plans in which the branches represent contingencies that may arise during execution. As the agent follows the plan, it evaluates the conditions on each branch to determine the most appropriate path to follow in the structure, thus dictating its actions. Another popular plan generation approach for BDI involves making use of an Hierarchical
Task Network (HTN)~\cite{Nau1999}; a planning method that divides complex plans into simpler subplans in a hierarchical manner. For example, Sardi\~na et al.~\cite{Sardina2006} make use of HTNs to expand selected plans into subplans that achieve the agent's current goal, while also preparing it for anticipated future goals. This provides a "lookahead" capability, allowing the agent to behave proactively. Many other planning approaches have been proposed for BDI, including some based on maximizing utilities~\cite{Walczak2007}, and statistics~\cite{Schut2002,Simari2006}. Although many of these approaches are effective, they  primarily rely on creating plans manually, which is error-prone and time-consuming.

%OTHER PLANNING APPROACHES
Many epistemic planning approaches have been proposed that, although not specifically designed for BDI, could be adapted for such use. While several of these approaches focus on single-agent systems~\cite{Bienvenu2010,Vezina2023,Andersen2012}, several multi-agent approaches have been proposed as well. For example, Torre~{n}o et al.~\cite{Torreno2014} introduced a multi-agent planning framework called Forward Multi-Agent Planning (FMAP), which enables agents to collaboratively explore the planning space and generate partial-order plans for achieving shared goals. Their approach employs a novel heuristic search algorithm and incorporates a privacy model that allows each agent to preserve the confidentiality of its private knowledge.
In other work, Muise et al.\cite{Muise2015} propose a method where individual agents reason as if they were other agents in order to derive strategies that support cooperation or competition. This is achieved by algorithmically formulating the multi-agent space into a single-agent planning problem from each agent’s perspective. A similar approach has been proposed by Kominis and Geffner\cite{Kominis2015}.
Despite the variety of proposed multi-agent planning approaches that have been proposed, applications of ATL in multi-agent planning theory appears to be unexplored. This is true within the context of BDI, and within the broader context of planning theory.

Even when  plans are carefully generated to account for a wide range of possible scenarios, a BDI agent must commit to all the steps in a selected plan until it is completed. This is a major limitation of BDI, as there is no guarantee that a selected plan will remain relevant throughout its execution~\cite{Meneguzzi2013}.  To overcome this problem, a more modern approach to planning in BDI is to generate relevant plans automatically during execution. This not only reduces the need for manual plan creation, but also enables agents to construct specific plans at runtime. For instance, in the work by Silva et al.~\cite{DeSilva2009}, an agent dynamically generates new plans based on manually defined ones, its current beliefs, and its various goals. To do so, a tree is built from the manually inputted plans, representing a hybridized model of these plans. At runtime,  a custom algorithm  dynamically searches this tree based on unique combinations of  the agent's beliefs and goals to create new plans. A similar approach for using previously defined plans to generate new ones at runtime was also proposed by Despouys and Ingrand~\cite{Despouys2000}. In another work by M\'ora et al.~\cite{mora1999bdi}, an event calculus was developed to model individual agents and their actions.  Through the use of algebraic methods, agents can formally verify at runtime whether certain goals are attainable and identify the actions needed to achieve them. Another modeling approach for enabling BDI agents to generate plans at runtime was also proposed by Meneguzzi and Luck~\cite{Meneguzzi2008}.

In this work, we inspire ourselves from these runtime approaches to generate plans automatically using ATL. Compared to existing planning approaches, ATL naturally supports multi-agent systems, allowing the generation of high-quality plans that account for both competition and cooperation among agents. As the use of ATL to generate plans for BDI agents remains unexplored, the work presented in this paper is novel. 
% End Section

\section{Motivating ATL-Supported BDI Agents}
\label{sec:ATLBDI}
% Begin Section
This section motivates how ATL can guide BDI agents' actions at runtime.
Although the content of this section is primarily conceptual, it lays the foundation for the remainder of this work. 
% We begin this section by adapting the definitions of beliefs, desires, and intentions presented in Section~\ref{sec:background} to align with the terminology of ATL. As a reminder, ATL formulas are applied to CGM or CGMII that capture the possible states of the environment, the agents in the system, and the resulting states from the actions of agents. In the case of CGMII, it also captures the uncertainty of each respective agent. 
% From the content presented in Section~\ref{sec:background}, 
We can argue that a CGMII can be seen as a representation of how the agents perceive their world, serving as a snapshot of the agents' certainty and uncertainty about the variables in their environment. 
% It also specifies the agent's beliefs about the certainty and uncertainty of other agents.  
% With this idea in mind, we re-align the definitions of beliefs, desires, and intentions for ATL-based BDI as follows:
In the context of BDI, this means the agents' beliefs could be represented using such a model, with their desires expressed as ATL formulas, and the strategies derived from these formulas representing their intentions.

% \begin{itemize}
%     \item Beliefs: A set of values believed to be true by an agent, including a CGMII model that represents the agent's own certainty and uncertainty for the values in the system, and its perceptions of the certainty and uncertainty of other agents.
%     \item Desires: A set of ATL formulas for which the agent exhibits ``pro-attitudes''. 
%     \item Intentions: Describes a commitment by the agent resulting from the strategies prescribed by the desired ATL formulas applied to the agent's CGMII belief model.
% \end{itemize}

In practice, this means that the agents would continuously construct a CGMII model as they operate. Realistically, any component of the CGMII model, such as the  agents, the states, and the propositions, could change over time, regardless of agent actions, potentially requiring very different CGMII models to be constructed as the agents operate. More interestingly, in an environment where only the agents' beliefs about the environmental variables evolve over time, which we define as an \textit{epistemic} environment, the only component in this model that would change  is each agent's indistinguishability (i.e., $\{\sim  i\}_{i\in \Sigma}$). This is because as the agents operate in their environment and perceive their world, they can gain knowledge about the values of the propositions in the environment. Under the assumption that the environment's states are defined by these propositions, this gain in knowledge may eliminate, or even create, indistinguishability for the individual agents.

\section{ATLantis}
\label{sec:ATLantis}
% Begin Section
In epistemic environments, the concept presented in Section~\ref{sec:ATLBDI} requires a CGMII model to be updated after each agent's perception. While this is feasible, there is a computational cost associated with updating a CGMII model and verifying an ATL formula  after each perception. As the computational cost of verifying an ATL formula on a CGMII is PSPACE-complete~\cite{Vester2013}, the approach is prone to real-time performance delays when used practically.

As previously described, for epistemic environments, the only element of the CGMII model that will change is  $\{\sim  i\}_{i\in \Sigma}$. Therefore, we can avoid the computational overhead involved in continuously updating a CGMII model and  verifying ATL formulas in real-time by evaluating all possible agent desires (represented as ATL formulas) across all possible combinations of values in ${\sim i}_{i \in \Sigma}$ on a given CGMII model.  In this section, we present a tool called ATLantis designed specifically to perform this task. 
% The tool outputs AgentSpeak plans with pre-conditions set to an agent's desire and a possible permutation of $\{\sim  i\}_{i\in \Sigma}$ (representing the agent's uncertainty). As a result, each plan specifies the actions an agent must take to achieve its desire, given its uncertainty of the environment. 
By pre-generating all possible strategies, we avoid the real-time performance overhead of analyzing ATL formulas during execution. Note that ATLantis is publicly available at: \url{https://github.com/DylanLeveille/ATLantis}.

\begin{figure}[ht!]
    \centering
    \centerline{\includegraphics[width=0.85\textwidth]{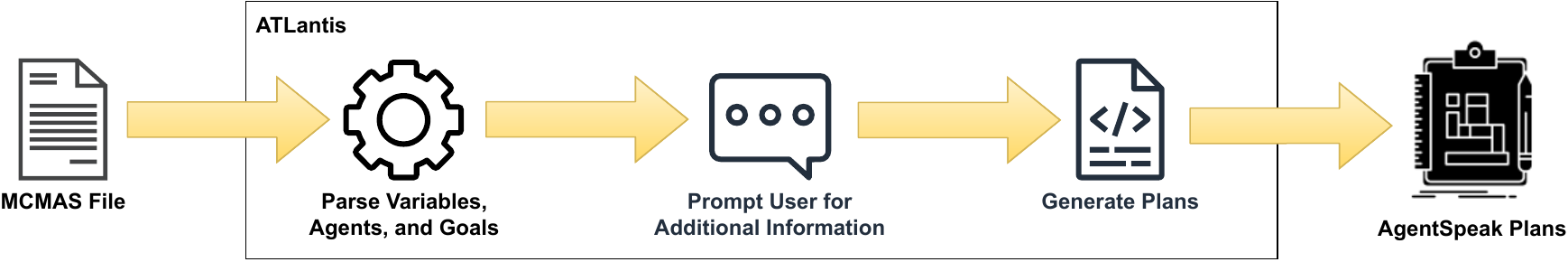}}
    \caption{An overview of ATLantis's design and functionality}
    \label{fig:ATLantisOverview}
\end{figure}

\begin{figure}[ht!]
    \centering
    \centerline{\includegraphics[width=0.2\textwidth]{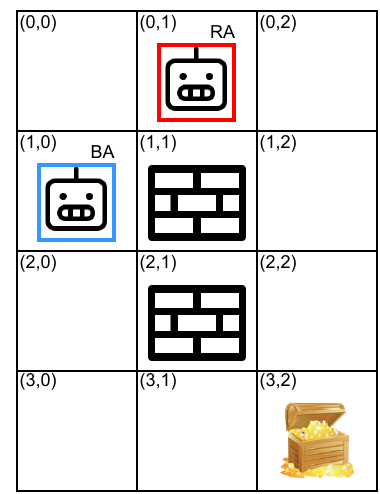}}
    \caption{The \goldseeker game}
    \label{fig:toySystem}
\end{figure}

\begin{figure}[ht!]
\centering
\begin{lstlisting}[language=ISPL]
Agent Environment
    Vars:
        rowBA : 0..3;
        columnBA : 0..2;
        rowRA : 0..3;
        columnRA : 0..2;
        treasureMined: boolean; 
    end Vars
    Evolution:
        rowBA = rowBA + 1 and rowRA = rowRA + 1 if ...
    end Evolution
end Agent

Agent BA
    Vars:
        mined: boolean;
    end Vars
    Actions = {right, left, up, down, mine};
    Evolution:
        mined = true if BA.Action = mine;
    end Evolution
end Agent

Agent RA
...
end Agent

Evaluation
    BATreasure if  Environment.rowBA=3 and  Environment.columnBA=2;
    RATreasure if  Environment.rowRA=3 and  Environment.columnRA=2;
    BAMined if  BA.mined = true;
    RAMined if  RA.mined = true;
    treasureTaken if Environment.treasureMined = true; 
end Evaluation

Groups
    g1 = {BA,RA};
end Groups

Formulae
    <g1> F(BATreasure and RATreasure and !BAMined and !RAMined and (<g1> X(treasureTaken)));
end Formulae
\end{lstlisting}
\caption{The MCMAS file for the \goldseeker game}
\label{fig:isplToySystem}
\end{figure}

An overview of ATLantis’s design and functionality can be seen in Figure~\ref{fig:ATLantisOverview}. The remainder of this section describes the components of this design. To illustrate the different steps in ATLantis, we will apply them to a small game called \goldseeker. This game can be seen in Figure~\ref{fig:toySystem}. There are two agents in this game, denoted as \texttt{BA} and \texttt{RA}. The agents are in cooperation, and wish to mine the treasure located at (3,2). To obtain the treasure, both agents must mine it while at its location. Although the agents’ initial positions are randomized, the treasure is always located at the same coordinate.  As a result, each agent is uncertain of their own initial position, as well as the initial position of the other agent. However, each agent has full visibility into the row and column they occupy, and can use this information to help infer their own position, and the position of the other agent.
Agents can move \textit{down}, \textit{up}, \textit{left}, or \textit{right}.  Agents cannot move beyond the coordinates, or through obstacles. For example, should an agent be at coordinate (1,0), moving \textit{right} or \textit{left} will keep them at this position.  Lastly, each agent is allowed to mine only once during the game.
Since both agents have individual uncertainties about their world and wish to cooperate towards a shared goal, this system presents an ideal use case for both BDI and ATL. 
% We will apply ATLantis to generate AgentSpeak plans for agent \texttt{BA} in this game. 

\subsection{MCMAS File}
MCMAS is a popular model checking toolkit for temporal logic and ATL~\cite{Lomuscio2017}. 
It allows ATL formulas to be verified against a CGMII model, and outputs a corresponding strategy if one exists. 
As input, MCMAS consumes an Interpreted Systems Programming Language (ISPL) file, which is internally converted to a CGMII model. 
% This input is specified using  . 
As ATLantis makes use of MCMAS, we adopt this style as the input format for ATLantis. In ATLantis, we refer to this file as the MCMAS file. An excerpt of the MCMAS file for the \goldseeker game is shown in Figure~\ref{fig:isplToySystem}. In this file, variables of the environment, and their evolution (i.e., how they change from the actions of agents), is specified in the \texttt{Environment} agent. Note that all specified variables are propositions. Integer variables, such as \textit{rowBA}, can only hold a single integer value. While not shown in this excerpt, the \textit{treasureMined} variable is set to \textit{true} only when both agents perform the mine action at location (3,2). Agents \texttt{BA} and \texttt{RA} are specified below and have identical specifications. 
% Of particular interest is the \textit{Lobsvar} value, which specifies the variables whose values are known by the agent. In other words, it is used to specify $\{\sim  i\}_{i\in \Sigma}$. As input to ATLantis, this value must be set to an empty set. 
Lastly, 
% as ATLantis is being applied from the perspective of agent \texttt{BA}, 
we specify an ATL formula that represents the agents' desire for eventually obtaining the treasure.  Since each agent can mine only once, we specify a formula that asks whether there exists a strategy involving cooperation between the two agents to reach the treasure without mining, followed by an action that allows them to mine it.  The complete MCMAS file for this system can be viewed in the ATLantis repository.  

\subsection{Parse Variables, Agents, and Goals}
ATLantis is a \texttt{Python}~\cite{van1995python} program that can be started from the command line. Once started, the user is prompted to select an MCMAS file through a file picker. The variables of the environment and of individual agents, the agents themselves, and the goals specified will be parsed from this file.

\subsection{Prompt User for Additional Information}

\begin{figure}[ht!]
    \centering
    \centerline{\includegraphics[width=\textwidth]{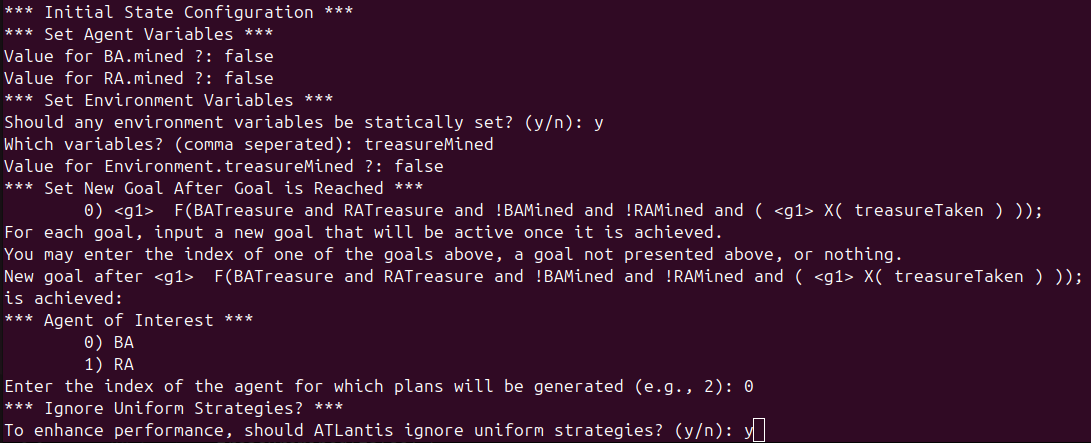}}
    \caption{Example ATLantis prompts for the \goldseeker game}
    \label{fig:ATLantisUserPrompt}
\end{figure}

Once the file is parsed, the user will be prompted for additional information. Example prompts for \goldseeker can be seen in Figure~\ref{fig:ATLantisUserPrompt}. In ATLantis, agent variables must be set to a fixed initial value. This is because agent variables are independent of environment variables, and should always be initialized to the same default values, regardless of the environment state.  For example, in Figure~\ref{fig:ATLantisUserPrompt}, the value \textit{false} is assigned to the agents’ internal variables, indicating that no agent has mined when initially placed in the game.
After setting the agent variables, ATLantis will prompt the user whether any environment variables should be statically configured. As shown in Figure~\ref{fig:ATLantisUserPrompt}, the \textit{treasureMined} variable is set to \textit{false} as the treasure will never be mined at the start of the game. Then, the goals to be set after achieving each ATL goal are specified. In this example, no subsequent goal is provided to the ATL goal specified in the MCMAS file. This implies that no future goal should be set after obtaining the treasure. 

MCMAS allows the indistinguishability of at most one agent to be specified; all other agents are assumed to have complete certainty. The implications of this limitation are discussed in Section~\ref{sec:discussion}. As a result, ATLantis prompts the user to select the agent that will exhibit uncertainty in its environment. This is the agent that ATLantis will generate AgentSpeak plans for. In this example, agent \texttt{BA} is selected. Lastly,  the user is prompted whether ATLantis should ignore uniform strategies when verifying ATL formulas. Since uniform strategies are unlikely in games where agents have freedom of movement, we ignore them for \goldseeker. To support this claim, in \goldseeker, if both agents are at the treasure, they could both mine directly, move \textit{right} and then \textit{mine}, or move \textit{right} twice and then \textit{mine}, etc.

\subsection{Generate Plans}
After the prompts are processed, ATLantis will generate the AgentSpeak plans for the agent previously selected. Let $G$ be the set of ATL goals specified in the input file, and $V$ be the set of environment variables. The algorithm used for generating these plans is shown in Algorithm~\ref{alg:planGenerator}.

\begin{algorithm}[t!]
\caption{ATLantis algorithm for generating AgentSpeak plans}\label{alg:planGenerator}
\begin{algorithmic}[1]
\Procedure{GeneratePlans}{$G$, $V$}
% \If{cost($\textit{optControls}$) $\leq$ $\textit{budget}$}
% \State \Return $\textit{optControls}$
% \EndIf
\State \textit{plans}  $\gets \emptyset$
% \State \textit{numAgents}  $\gets |A|$
\State \textit{agentVariableSubsets}  $\gets \powerset(V) $
%  \For{$i=1; i<\textit{numAgents};\textit{i++}$}
%     \State \textit{allAgentVariablePermuations}  $\gets \textit{allAgentVariablePermuations} \times \mathcal{P}(V)$
% \EndFor
\For{$g$ \texttt{in} $G$}
    \State \texttt{Set MCMAS Formulae as} $\; g$ 
    \For{$\textit{agentVariableSubset}$ \texttt{in} $\textit{agentVariableSubsets}$}
    % \For{$i=0; i<\textit{numAgents};\textit{i++}$}
    %     \State $\textit{agentVariables} \gets \textit{agentVariablePermutations[i]}$
    %     \State \texttt{Set MCMAS Lobsvar for agent} $A_{i}$ \texttt{as} \textit{agentVariables}
    % \EndFor
    \State $\textit{knownVars} \gets \textit{agentVariableSubset}$
    \State $\textit{unknownVars} \gets V -\textit{knownVars} $
    \State \textit{n}  $\gets |\textit{knownVars}| $
    \State \textit{n'}  $\gets |\textit{unknownVars}| $
    \State \textit{knownPossValues}  $\gets \textit{knownVars}[0]\textit{.values} \times \textit{knownVars}[1]\textit{.values} \times ... \times \textit{knownVars}[n]\textit{.values} $
    \State \textit{unknownPossValues}  $\gets (\powerset(\textit{unknownVars}[0]\textit{.values}) - \emptyset )\times (\powerset (\textit{unknownVars}[1]\textit{.values}) - \emptyset ) \times ... \times (\powerset(\textit{unknownVars}[n']\textit{.values}) - \emptyset ) $
    \For{$\textit{knownPossValue}$ \texttt{in}  $\textit{knownPossValues}$}
        \For{$\textit{unknownPossValue}$ \texttt{in}  $\textit{unknownPossValues}$}
              \State \texttt{Set MCMAS InitStates \textbackslash w} \; \textit{knownPossValue, unknownPossValue} 
            \State $\textit{strategy} \gets$ \Call{FindUniformStrategy}{$\texttt{MCMAS}$}
                \If{ \texttt{NOT} $(\Call{Exists}{\textit{strategy}} )$}
                \State $\textit{strategy} \gets$ \Call{FindStrategy}{$\texttt{MCMAS}$}
                    \If{ \texttt{NOT} $(\Call{Exists}{\textit{strategy}} )$}
                    \State $\textit{strategy} \gets$ \Call{Prev}{\textit{knownPossValue}, \textit{unknownPossValue}}
                    \EndIf
                \EndIf
            
            \State $\textit{plans.add}(\textit{strategy})$
        \EndFor
    \EndFor
    
\EndFor
\EndFor
\State \Return \textit{plans}
\EndProcedure
\end{algorithmic}
\end{algorithm}

As the agent provided to ATLantis can know the values of any number of variables in the system, plans must be generated for every possible combination of known variables. The algorithm keeps track of this in the \textit{agentVariableSubsets} variable, which is initialized to the powerset of $V$ (line 3). This results in a set containing all subsets of environment variables that could be known by the agent. 

Next, the algorithm iterates through each goal in $G$, and sets the  formulae accordingly for each goal in MCMAS (line 5). The algorithm then iterates through the subsets of environment variables that could by known by the agent, and stores this value in \textit{knownVars}  (line 7). The unknown variables are subsequently computed (line 8). The algorithm then computes all possible subsets of the values that each of the known variables can have (line 11). As these variables are known by the agent, each  must have one unique value. Similarly, all possible subsets of the values that each of the unknown variables can have is computed (line 12). Unlike the known variables, the unknown variables must possess a range of possible values. This range is computed by performing a powerset on the possible values for each variable, and removing the empty set. While not explicitly shown in Algorithm~\ref{alg:planGenerator}, singleton sets are also removed from each powerset (as only known variables can possess a single possible value). 

The algorithm then iterates over the possible values of both known and unknown variables (lines 13-14), and sets the initial state in MCMAS (line 15). This initial state corresponds to $\{\sim  i\}_{i\in \Sigma}$ in the CGMII model created by MCMAS.
% , as it represents the uncertainty of the agent. 
MCMAS then attempts to find a uniform strategy for the agent (line 16). Note that ATLantis will not attempt to find a uniform strategy if the user has opted to ignore them. If no uniform strategy exists, or if the user chooses to ignore them, a non-uniform strategy is computed instead (line 18). 

If the ATL formula does not hold under the given constraints, no non-uniform strategy will be found. As a result, the algorithm attempts to retrieve a strategy that was previously computed by refining the possible values of the unknown variables. To do so, the algorithm statically assigns values to the unknown variables based on their possible ranges. It then attempts to identify a strategy that was previously found with this configuration.
For example, in \goldseeker, let's assume no strategy was found when the unknown variables were \textit{rowRA} and \textit{columnRA}, and where the range of the values is either 0 or 1. The algorithm would iteratively assigns concrete values to these variables (e.g., \textit{rowBA} = 0 and \textit{columnBA} = 1) and check whether a previously computed strategy exists under those conditions. If no strategy can be found for any concrete assignment of the unknown variables, this indicates that the agent is unable to achieve its goal under any configuration within the range of possible values. This logic is encapsulated in the \Call{prev}{} function. If no strategy can be found by refining the uncertainty of variables in the \Call{prev}{} function, this implies that the goal is unachievable given the agent’s uncertainty.

The computational complexity of the algorithm can be calculated as follows. Let $l$ be the maximum length of the range of possible values among all variables. Note that there are $|G|$ goals, and $ 2^{|V|}$ subsets in $\textit{agentVariableSubsets}$. Also note that the worst case length of $\textit{unknownPossValues}$ surpasses the worst case length of  $\textit{knownPossValues}$, and that the worst case length of $\textit{unknownPossValues}$ is $2^{l} \cdot |V|$. As a result, disregarding the complexity of MCMAS and of the \Call{prev}{} function, the computational complexity of this algorithm is $O(|G| \cdot 2^{|V|} \cdot 2^{l} \cdot |V|) = O(|G| \cdot |V| \cdot 2^{l|V|})$. We ignore the performance cost of MCMAS and the \Call{prev}{} function in this calculation, as the computed efficiency is intended to reflect the number of iterations performed by the algorithm to find strategies.

\subsection{AgentSpeak Plans}

When the algorithm completes, ATLantis will output the generated AgentSpeak plans for the agent. Each plan has its pre-conditions set to a unique combination of known variable values, and possible ranges for the unknown variables. As a result, the pre-conditions of each plan generated are mutually exclusive. This ensures that for any given combination of the agent’s certainties and uncertainties, only one plan is selected. For agent \texttt{BA}, a total of 11025 plans were generated. One of these plans can be seen in Figure~\ref{fig:AgentSpeakToySystem}.  For example, in Figure~\ref{fig:AgentSpeakToySystem},  the pre-conditions are configured for the case where agent \texttt{BA} knows with certainty that it is in column 0 and row 2, while being uncertain about agent \texttt{RA}’s exact position. However, it believes that agent \texttt{RA} is in either column 1 or 2, and in row 0, 1, 2, or 3. Also note that the previously specified fixed variable is included in the pre-conditions of all generated plans. When a plan is selected, the Jason-specific command \texttt{.drop_all_intentions} is invoked to clear all previously selected plans.  This is because selecting a new plan reflects a change in the agent’s beliefs, making any previously selected plan obsolete. 
% In fact, as noted in Section~\ref{sec:relatedwork}, all steps of a plan must be completed once selected. This command ensures that outdated plans are discarded. 
After dropping any previously selected plan, the agent will proceed to execute the actions in the newly selected plan. As no new goal was specified upon obtaining the treasure, the plan terminates with the Jason command \textit{true}, which serves as a ``no-action'' operation. The plans for agents \texttt{BA} and \texttt{RA} were both generated using ATLantis, and are available in the ATLantis repository.

\begin{figure}[ht!]
\centering
\begin{lstlisting}[language=AgentSpeak]
+!gettreasure:
	treasuremined(false) & columnba(0) & rowba(2) & poss(columnra(1)) &
	poss(columnra(2)) & poss(rowra(0)) & poss(rowra(1)) & poss(rowra(2)) &
	poss(rowra(3))
	<-
	.drop_all_intentions; down; right; right; mine; true.
\end{lstlisting}
\caption{A sample AgentSpeak plan generated by ATLantis for agent \texttt{BA}}
\label{fig:AgentSpeakToySystem}
\end{figure}

% \mynote{0}{Great in theory, but we need plans. Bratman says plans crucial.  Plans could be generated dynamically/runtime as the ATL model is created, or all pre-execution. In static environment, as said earlier, tilda decides ATL model. As a result tilda serves as pre-consitions of plans.  ATLantis can generate plans for this ATL-Based BDI approach for static environments.}
% End Section

\section{Applying ATLantis Plans at Runtime}
\label{sec:uncertainty}
% Begin Section
In this section, we explain and demonstrate how the plans generated by ATLantis can be applied during an agent's runtime. In an epistemic environment, to use the plans generated by ATLantis, an agent would only need to track its uncertainty with respect to the propositions in its environment to select applicable plans. As these propositions represent epistemic values believed by an agent, an epistemic model, such as Public Announcement Logic (PAL)~\cite{VanDrimmelen2023PAL} or Dynamic Epistemic Logic (DEL)~\cite{Baltag2016} could be used to hold each agent's uncertainty. With such a model, as the agent perceives its environment, the agent's epistemic model can be updated. A plan whose preconditions align with the updated epistemic model can then be selected. In fact, this approach has already been proposed by Vezina et al.~\cite{Vezina2023}. We will adopt a similar approach to demonstrate how the plans generated by ATLantis would be selected by the agents at runtime in the \goldseeker game. The following subsections will demonstrate the belief updates and the corresponding selected plan after each agent perception for each agent. The initial position of both agents will be the configuration shown in Figure~\ref{fig:toySystem}. As a reminder, each agent has visibility into the row and column they occupy, and can use their perceptions to reduce uncertainty about their position. Specifically, they can use the locations of obstacles to help infer their position.

\subsection{Perception 1}

\begin{figure}[H]
    \centering
    \centerline{\includegraphics[width=0.7\textwidth]{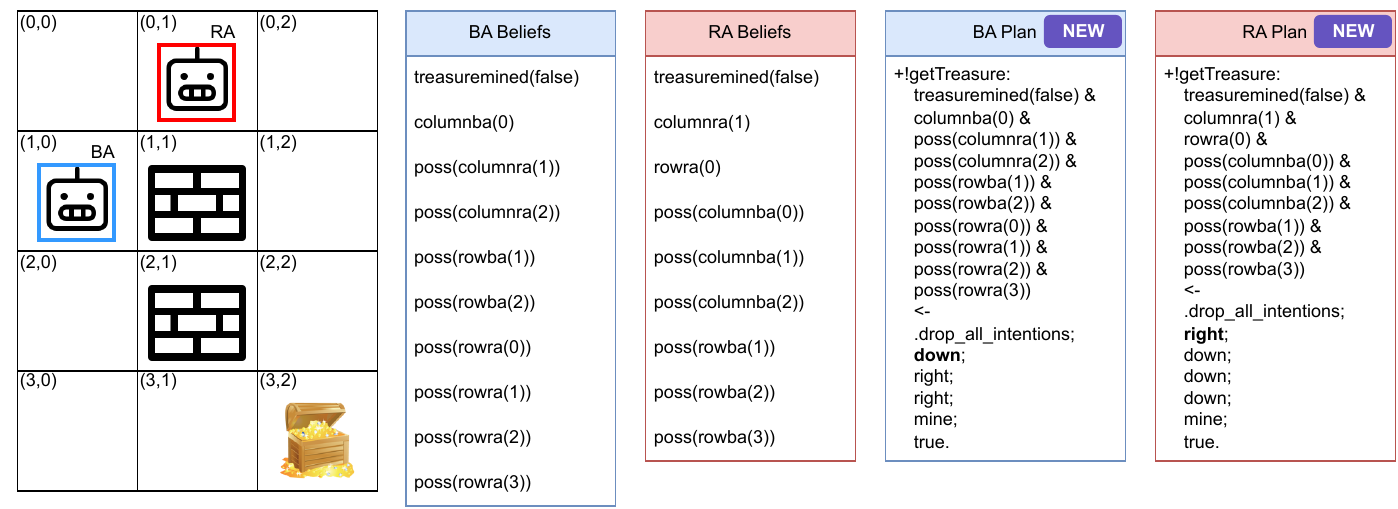}}
    \caption{Resulting game state after perception 1}
    \label{fig:percept1}
\end{figure}

Upon starting the game, agent \texttt{RA} immediately knows its position. This is because there is only one position on the map with an obstacle directly south of the player. Since each player has visibility into their row and column, agent \texttt{RA} can also infer that agent \texttt{BA} is not in row 0, but could be in any of the other rows. However, due to an obstacle blocking its vertical view, \texttt{RA} assumes that agent \texttt{BA} is in any column. As for agent \texttt{BA}, there are two possible positions where an obstacle exists to the east of the player. Therefore, agent \texttt{BA} can assume it is at either coordinate (1,0) or (2,0). As a result, it knows it must be in column 0, but that its row could be either 1 or 2. Similarly, \texttt{BA} knows that agent \texttt{RA} is not in column 0 but could be in any row.
Based on these beliefs, each agent selects the plan corresponding to its current knowledge and applies the first action from its respective plan.

\subsection{Perception 2}

\begin{figure}[H]
    \centering
    \centerline{\includegraphics[width=0.7\textwidth]{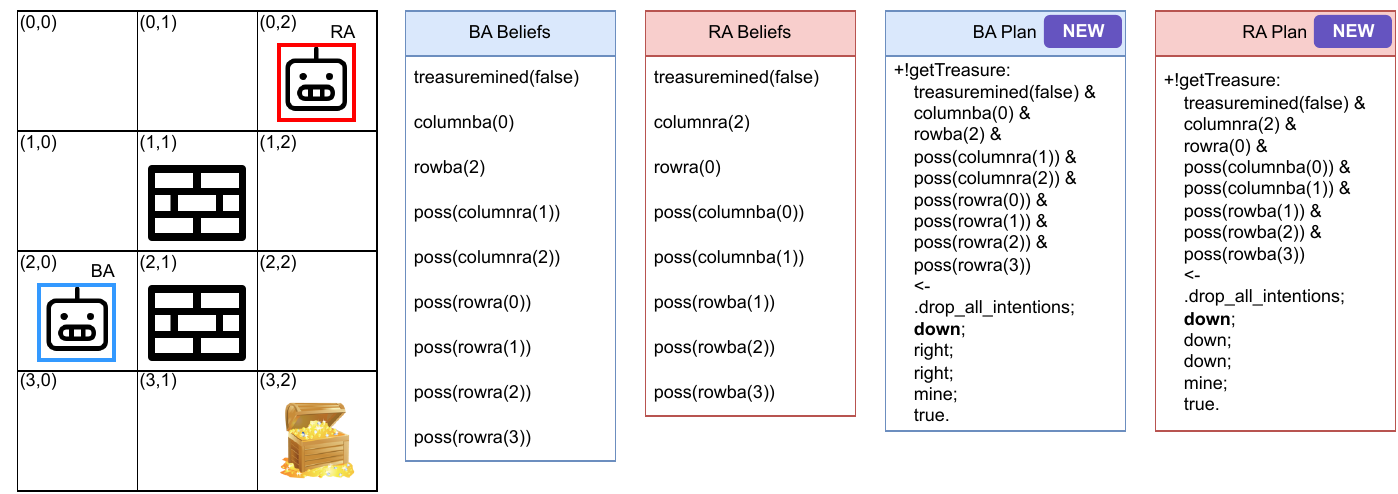}}
    \caption{Resulting game state after perception 2}
    \label{fig:percept2}
\end{figure}

In the next perception, agent \texttt{RA} remains certain of its position (now at (0,2)), since knowing its previous position allows it to infer its new location after moving. From (0,2), agent \texttt{RA} has full visibility of row 0 and column 2. It does not see agent \texttt{BA} and thus assumes that \texttt{BA} must be located in the other rows and columns.
As for agent \texttt{BA}, it now also knows its position. As there are two possible locations where an obstacle is present to the east, after moving \textit{down} and still observing an obstacle to the east, \texttt{BA} can conclude it must be at (2,0). As before, \texttt{BA} infers that \texttt{RA} is not in column 0 but could be in any row.
Since the agents' beliefs have changed, a new plan must be selected. The previous plans are abandoned, and each agent applies the first action from their newly selected plan.

\subsection{Perception 3}

\begin{figure}[H]
    \centering
    \centerline{\includegraphics[width=0.7\textwidth]{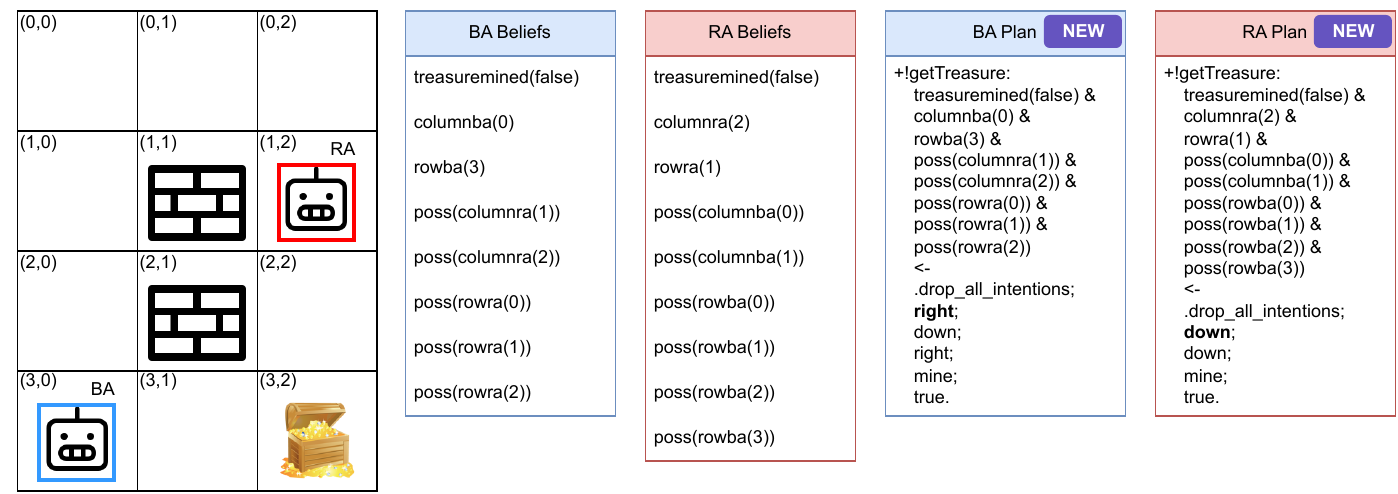}}
    \caption{Resulting game state after perception 3}
    \label{fig:percept3}
\end{figure}

After the third perception, both agents are again certain of their own positions. Agent \texttt{RA} can conclude that \texttt{BA} is not in column 2 but can no longer infer \texttt{BA}'s row. Meanwhile, agent \texttt{BA} concludes that \texttt{RA} cannot be in row 3 or column 0. Since the agents' beliefs have changed once more, the previous plans are abandoned, new plans are selected, and the first action of each new plan is executed.

\subsection{Perception 4}

\begin{figure}[H]
    \centering
    \centerline{\includegraphics[width=0.7\textwidth]{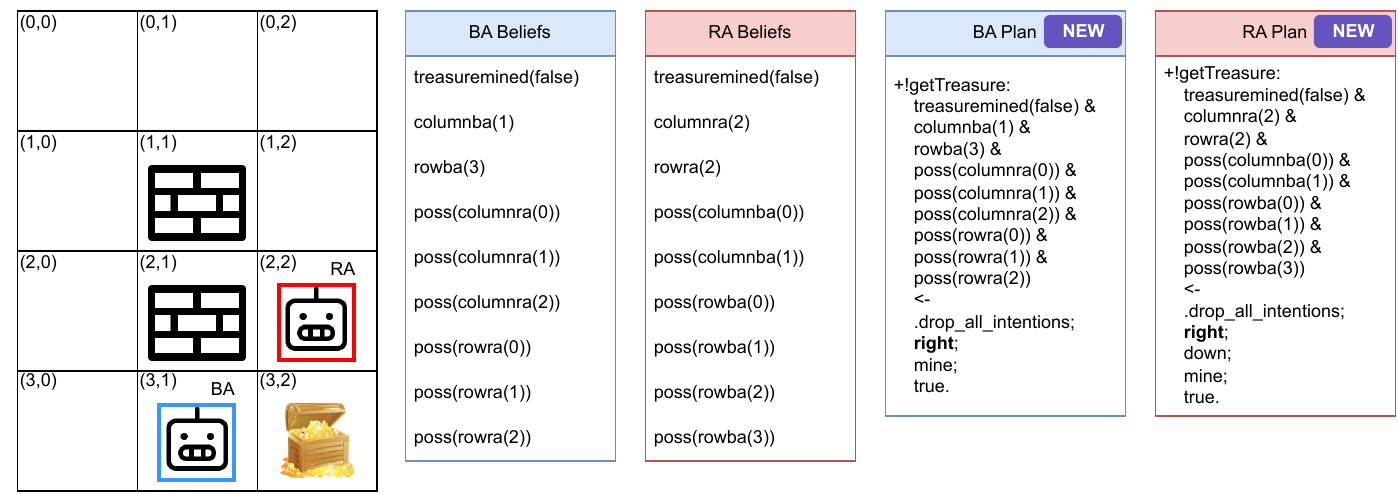}}
    \caption{Resulting game state after perception 4}
    \label{fig:percept4}
\end{figure}

After the fourth perception, both agents remain certain of their own positions and are nearing the treasure. As before, agent \texttt{RA} can infer that agent \texttt{BA} is not in column 2 but cannot determine \texttt{BA}’s row. Similarly, agent \texttt{BA} knows that \texttt{RA} is not in row 3 but cannot conclude \texttt{RA}’s column. Both agents select new applicable plans and apply the first action of their respective plans.
However, we observe something interesting with the plan selected by \texttt{RA}. Until now, all plans have resulted in a change to the agents' positions. However, \texttt{RA}’s current plan instructs it to move \textit{right}, thus making it stay at the same location. This occurs because uniform strategies do not exist for the goal provided in \goldseeker. As a result, many possible strategies can be generated by MCMAS through ATLantis. In this particular case, the strategy provided does not guide the agent directly to the treasure. However, it is still valid according to the agent’s belief about the world, and does intend to lead the agent to the treasure, and to mine it.
Due to space limitations, we omit the remaining perception cycles. However, in the remaining cycles, the agents eventually learn each other’s positions and will mine the treasure together.

% \subsection{Perception 5}

% \begin{figure}[H]
%     \centering
%     \centerline{\includegraphics[width=0.7\textwidth]{Figures/epistemicUpdates-iteration5.drawio.pdf}}
%     \caption{Resulting system state after perception 5}
%     \label{fig:percept5}
% \end{figure}
% End Section

\section{Discussion}
\label{sec:discussion}
% Begin Section
%Limitation 1: MCMAS not great and unssuported. More importantly, cannot get strategies for formulas with G and U. 

%%%%%%%%%%%%%%%%%%%%%%%%%%%%%%%%%%

%*Limitation 3: Correctness of CGMII model. Super important for both cooperation and competition. If model is wrong, startegies could be very wrong as well. Stratgies are deterministic; may not hold for non-deterministic worlds
With ATLantis,  plans for BDI agents can be automatically generated from provided ATL goals, thus creating plans that allow for  competitive and cooperative behaviour.   Since ATLantis generates plans prior to runtime, we avoid the performance overhead with dynamically creating plans during execution. Lastly, since ATLantis generates plans automatically, the need to manually create plans is eliminated. 

%10 lines
A possible limitation with the plans generated by ATLantis is that they depend heavily on the correctness of the provided MCMAS file (i.e., the CGMII model representing the environment). Small mistakes in the provided MCMAS file can lead to plans that violate intended competitive or cooperative behaviours. While this issue can be avoided by providing a correct MCMAS input file, any properties of the environment or agents that are potentially unknown, and therefore not captured in this file, may result in undesirable behaviours. Related to this point, as CGMII models  are deterministic, the plans generated by ATLantis may not work for non-deterministic environments. In other words, the plans generated only work for epistemic environments. ATLantis is unable to generate plans for environments with continuously emerging variables and new agent actions. Unfortunately, these issues present a greater limitation with ATL as a whole. However, in correctly specified deterministic environments, the plans generated by ATLantis will allow the agents to exhibit the intended behaviours for the given goals.

%Limitattion 2: MCMAS can only put uncertainty of one agent. not great
%Limitation 6: In ATL, a model can prescirbe uncertainties of multiple agents. Like a grouping of uncertainties of both. Practicality, this would to create a model from individual beliefs of each agent. Strategies would have to be transfered to indifvidiual agents. Not realusitc. 
Another limitation with ATLantis is that it only allows the uncertainty of a single agent to be specified, while all other agents are assumed to have complete certainty. This constraint is inherited from MCMAS itself. As a result, plans are generated under the assumption that other agents have complete knowledge of the environment's variables. While this may seem unrealistic, an alternative solution, which is to generate plans that require  an agent to know the variables known by all other agents, is even less realistic. In fact, this could only be achieved practically if all agents communicated their uncertainty to each other at runtime. We cannot expect all agents to do this, especially if they are in competition. Given this argument, it is much more realistic to assume that other agents  possess complete certainty, thus implying they act confidently throughout the system. While it may be reasonable to generate plans that assume full certainty for some agents and partial certainty for others, this is not supported by MCMAS, and therefore not supported by ATLantis.
% as it at least allows the generation of plans in which the other agents are assumed to act with full confidence. 

%Limitation 4: Not in real-time, so we are OK. TOok only 7 minutes to generate all plans for \goldseeker.

%Limitation 5: If no uniform strategy, how to know we have the 'best' one...This is difficult. Would require another tool beside MCMAS, since we can't control its results. 
Another more minor limitation includes the handling of situations where many non-uniform strategies are found by MCMAS for a given ATL goal. In such cases, only one strategy is selected by MCMAS and returned to ATLantis.  Ideally, it would be preferable to compare these strategies to determine the ``best'' one. However, as this decision is handled internally by MCMAS, it is not something that can be controlled by ATLantis. More importantly, evaluating which non-uniform strategy is best would require defining metrics, which could vary significantly between environments.
Another minor limitation is ensuring that cooperative agents agree on the same ATL goal before pursuing it. This issue falls outside the scope of both ATL and BDI, and is more broadly related to the general problem of agent cooperation. However, in practice, agents could simply communicate and agree on cooperative ATL goals before acting on them.
Lastly, as a final limitation, as found in Section~\ref{sec:ATLantis}, while ATLantis runs in exponential time, results can be generated quickly. In fact, the results for the \goldseeker game were generated in just under five minutes. However, since ATLantis has only been used to generate plans for the \goldseeker game, its computational performance on larger systems remains untested.  

% End Section

\section{Conclusions and Future Work}
\label{sec:conclusion}
% Begin Section
ATL is a powerful language for specifying goals in multi-agent systems that involve both competition and cooperation. In this work, we present a tool called ATLantis which uses ATL to automatically generate plans for BDI agents. As input, ATLantis consumes an MCMAS file representing a CGMII model of the environment and the agents. To demonstrate the effectiveness of the tool, we apply ATLantis and its generated plans to an illustrative game that requires cooperation to achieve a shared goal. 

In future work, we aim to replace the ATL model checker used by ATLantis (i.e., MCMAS) as many of ATLantis's limitations, such as the inability to specify the uncertainty of multiple agents, and the lack of control over the selection of non-uniform strategies, are due to constraints in  MCMAS. Since no other ATL model checkers are widely used in the literature, this replacement may require the development of a custom ATL model checker. Additionally, while ATLantis has been shown to run in exponential time, its practical performance limitations have not been evaluated experimentally. Therefore, in future work, we will evaluate ATLantis' performance by empirically measuring its efficiency in generating plans for multiple complex systems.
% End Section

% Bibliography
\bibliographystyle{eptcs}
\bibliography{GandALF2025}

\begin{thebibliography}{10}
\providecommand{\bibitemdeclare}[2]{}
\providecommand{\surnamestart}{}
\providecommand{\surnameend}{}
\providecommand{\urlprefix}{Available at }
\providecommand{\url}[1]{\texttt{#1}}
\providecommand{\href}[2]{\texttt{#2}}
\providecommand{\urlalt}[2]{\href{#1}{#2}}
\providecommand{\doi}[1]{doi:\urlalt{https://doi.org/#1}{#1}}
\providecommand{\eprint}[1]{arXiv:\urlalt{https://arxiv.org/abs/#1}{#1}}
\providecommand{\bibinfo}[2]{#2}

\bibitemdeclare{article}{Alur1998}
\bibitem{Alur1998}
\bibinfo{author}{Rajeev \surnamestart Alur\surnameend}, \bibinfo{author}{Thomas~A. \surnamestart Henzinger\surnameend} \& \bibinfo{author}{Orna \surnamestart Kupferman\surnameend} (\bibinfo{year}{1998}): \emph{\bibinfo{title}{{Alternating-time temporal logic*}}}.
\newblock {\slshape \bibinfo{journal}{Lecture Notes in Computer Science (including subseries Lecture Notes in Artificial Intelligence and Lecture Notes in Bioinformatics)}} \bibinfo{volume}{1536}, pp. \bibinfo{pages}{23--60}, \doi{10.1007/3-540-49213-5_2}.

\bibitemdeclare{article}{Andersen2012}
\bibitem{Andersen2012}
\bibinfo{author}{Mikkel~Birkegaard \surnamestart Andersen\surnameend}, \bibinfo{author}{Thomas \surnamestart Bolander\surnameend} \& \bibinfo{author}{Martin~Holm \surnamestart Jensen\surnameend} (\bibinfo{year}{2012}): \emph{\bibinfo{title}{{Conditional epistemic planning}}}.
\newblock {\slshape \bibinfo{journal}{Lecture Notes in Computer Science (including subseries Lecture Notes in Artificial Intelligence and Lecture Notes in Bioinformatics)}} \bibinfo{volume}{7519 LNAI}, pp. \bibinfo{pages}{94--106}, \doi{10.1007/978-3-642-33353-8_8}.

\bibitemdeclare{misc}{Baltag2016}
\bibitem{Baltag2016}
\bibinfo{author}{Alexandru \surnamestart Baltag\surnameend} \& \bibinfo{author}{Bryan \surnamestart Renne\surnameend} (\bibinfo{year}{2016}): \emph{\bibinfo{title}{{Dynamic Epistemic Logic}}}.
\newblock \bibinfo{howpublished}{\url{https://plato.stanford.edu/archives/win2016/entries/dynamic-epistemic/}}.

\bibitemdeclare{article}{Bienvenu2010}
\bibitem{Bienvenu2010}
\bibinfo{author}{Meghyn \surnamestart Bienvenu\surnameend}, \bibinfo{author}{H{\'{e}}l{\`{e}}ne \surnamestart Fargier\surnameend} \& \bibinfo{author}{Pierre \surnamestart Marquis\surnameend} (\bibinfo{year}{2010}): \emph{\bibinfo{title}{{Knowledge compilation in the modal logic S5}}}.
\newblock {\slshape \bibinfo{journal}{Proceedings of the National Conference on Artificial Intelligence}} \bibinfo{volume}{1}, pp. \bibinfo{pages}{261--266}, \doi{10.1609/aaai.v24i1.7587}.

\bibitemdeclare{book}{bordini2007programming}
\bibitem{bordini2007programming}
\bibinfo{author}{Rafael~H \surnamestart Bordini\surnameend}, \bibinfo{author}{Jomi~Fred \surnamestart H{\"u}bner\surnameend} \& \bibinfo{author}{Michael \surnamestart Wooldridge\surnameend} (\bibinfo{year}{2007}): \emph{\bibinfo{title}{Programming multi-agent systems in AgentSpeak using Jason}}.
\newblock \bibinfo{publisher}{John Wiley \& Sons}, \doi{10.1002/9780470061848}.

\bibitemdeclare{book}{Bratman1987-BRAIPA}
\bibitem{Bratman1987-BRAIPA}
\bibinfo{author}{Michael \surnamestart Bratman\surnameend} (\bibinfo{year}{1987}): \emph{\bibinfo{title}{Intention, Plans, and Practical Reason}}.
\newblock \bibinfo{publisher}{Cambridge, MA: Harvard University Press}, \bibinfo{address}{Cambridge}, \doi{10.2307/2185304}.

\bibitemdeclare{inproceedings}{davies2024event}
\bibitem{davies2024event}
\bibinfo{author}{Curtis \surnamestart Davies\surnameend} \& \bibinfo{author}{Babak \surnamestart Esfandiari\surnameend} (\bibinfo{year}{2024}): \emph{\bibinfo{title}{Event Sourcing in Jason: Event-Driven State Reconstruction for BDI Agents}}.
\newblock In: {\slshape \bibinfo{booktitle}{2024 IEEE International Conference on Agents (ICA)}}, \bibinfo{organization}{IEEE}, pp. \bibinfo{pages}{70--75}, \doi{10.1109/ica63002.2024.00023}.

\bibitemdeclare{article}{DeSilva2009}
\bibitem{DeSilva2009}
\bibinfo{author}{Lavindra \surnamestart {De Silva}\surnameend}, \bibinfo{author}{Sebastian \surnamestart Sardina\surnameend} \& \bibinfo{author}{Lin \surnamestart Padgham\surnameend} (\bibinfo{year}{2009}): \emph{\bibinfo{title}{{First principles planning in BDI systems}}}.
\newblock {\slshape \bibinfo{journal}{Proceedings of the International Joint Conference on Autonomous Agents and Multiagent Systems, AAMAS}} \bibinfo{volume}{2}, pp. \bibinfo{pages}{1006--1013}, \doi{10.1145/1558109.1558167}.

\bibitemdeclare{article}{Despouys2000}
\bibitem{Despouys2000}
\bibinfo{author}{Olivier \surnamestart Despouys\surnameend} \& \bibinfo{author}{Francois~Felix \surnamestart Ingrand\surnameend} (\bibinfo{year}{2000}): \emph{\bibinfo{title}{{Propice-plan: Toward a uniffed framework for planning and execution}}}.
\newblock {\slshape \bibinfo{journal}{Lecture Notes in Artificial Intelligence (Subseries of Lecture Notes in Computer Science)}} \bibinfo{volume}{1809}, pp. \bibinfo{pages}{278--293}, \doi{10.1007/10720246_22}.

\bibitemdeclare{article}{Georgeff1995}
\bibitem{Georgeff1995}
\bibinfo{author}{M.P. \surnamestart Georgeff\surnameend} \& \bibinfo{author}{A.S. \surnamestart Rao\surnameend} (\bibinfo{year}{1995}): \emph{\bibinfo{title}{{BDI agents: From theory to practice}}}.
\newblock {\slshape \bibinfo{journal}{Proceedings of the First International Conference on Multi-Agent Systems (ICMAS-95)}}, pp. \bibinfo{pages}{312--319}.

\bibitemdeclare{article}{Goranko2016}
\bibitem{Goranko2016}
\bibinfo{author}{Valentin \surnamestart Goranko\surnameend}, \bibinfo{author}{Antti \surnamestart Kuusisto\surnameend} \& \bibinfo{author}{Raine \surnamestart R{\"{o}}nnholm\surnameend} (\bibinfo{year}{2016}): \emph{\bibinfo{title}{{Game-theoretic semantics for alternating-time temporal logic}}}.
\newblock {\slshape \bibinfo{journal}{Proceedings of the International Joint Conference on Autonomous Agents and Multiagent Systems, AAMAS}}, pp. \bibinfo{pages}{671--679}, \doi{10.1145/3179998}.

\bibitemdeclare{article}{Kominis2015}
\bibitem{Kominis2015}
\bibinfo{author}{Filippos \surnamestart Kominis\surnameend} \& \bibinfo{author}{Hector \surnamestart Geffner\surnameend} (\bibinfo{year}{2015}): \emph{\bibinfo{title}{{Beliefs in multiagent planning: From one agent to many}}}.
\newblock {\slshape \bibinfo{journal}{Proceedings International Conference on Automated Planning and Scheduling, ICAPS}} \bibinfo{volume}{2015-January}, pp. \bibinfo{pages}{147--155}, \doi{10.1609/icaps.v25i1.13726}.

\bibitemdeclare{article}{Lomuscio2017}
\bibitem{Lomuscio2017}
\bibinfo{author}{Alessio \surnamestart Lomuscio\surnameend}, \bibinfo{author}{Hongyang \surnamestart Qu\surnameend} \& \bibinfo{author}{Franco \surnamestart Raimondi\surnameend} (\bibinfo{year}{2017}): \emph{\bibinfo{title}{{MCMAS: an open-source model checker for the verification of multi-agent systems}}}.
\newblock {\slshape \bibinfo{journal}{International Journal on Software Tools for Technology Transfer}} \bibinfo{volume}{19}(\bibinfo{number}{1}), pp. \bibinfo{pages}{9--30}, \doi{10.1007/s10009-015-0378-x}.

\bibitemdeclare{article}{Meneguzzi2013}
\bibitem{Meneguzzi2013}
\bibinfo{author}{Felipe \surnamestart Meneguzzi\surnameend} \& \bibinfo{author}{Lavindra \surnamestart {De Silva}\surnameend} (\bibinfo{year}{2013}): \emph{\bibinfo{title}{{Planning in BDI agents: A survey of the integration of planning algorithms and agent reasoning}}}.
\newblock {\slshape \bibinfo{journal}{Knowledge Engineering Review}} \bibinfo{volume}{30}(\bibinfo{number}{1}), pp. \bibinfo{pages}{1--44}, \doi{10.1017/S0269888913000337}.

\bibitemdeclare{article}{Meneguzzi2008}
\bibitem{Meneguzzi2008}
\bibinfo{author}{Felipe \surnamestart Meneguzzi\surnameend} \& \bibinfo{author}{Michael \surnamestart Luck\surnameend} (\bibinfo{year}{2008}): \emph{\bibinfo{title}{{Composing high-level plans for declarative agent programming}}}.
\newblock {\slshape \bibinfo{journal}{Lecture Notes in Computer Science (including subseries Lecture Notes in Artificial Intelligence and Lecture Notes in Bioinformatics)}} \bibinfo{volume}{4897 LNAI}, pp. \bibinfo{pages}{69--85}, \doi{10.1007/978-3-540-77564-5_5}.

\bibitemdeclare{misc}{Meuleau2012}
\bibitem{Meuleau2012}
\bibinfo{author}{Nicolas \surnamestart Meuleau\surnameend} \& \bibinfo{author}{David \surnamestart Smith\surnameend} (\bibinfo{year}{2012}): \emph{\bibinfo{title}{Optimal Limited Contingency Planning}}, \doi{10.5555/2100584.2100635}.

\bibitemdeclare{inproceedings}{mora1999bdi}
\bibitem{mora1999bdi}
\bibinfo{author}{Michael~C \surnamestart M{\'o}ra\surnameend}, \bibinfo{author}{Jos{\'e}~G \surnamestart Lopes\surnameend}, \bibinfo{author}{Rosa~M \surnamestart Viccariz\surnameend} \& \bibinfo{author}{Helder \surnamestart Coelho\surnameend} (\bibinfo{year}{1999}): \emph{\bibinfo{title}{BDI models and systems: Reducing the gap}}.
\newblock In: {\slshape \bibinfo{booktitle}{Intelligent Agents V: Agents Theories, Architectures, and Languages: 5th International Workshop, ATAL’98 Paris, France, July 4--7, 1998 Proceedings 5}}, \bibinfo{organization}{Springer}, pp. \bibinfo{pages}{11--27}, \doi{10.1007/3-540-49057-4_2}.

\bibitemdeclare{article}{Muise2015}
\bibitem{Muise2015}
\bibinfo{author}{Christian \surnamestart Muise\surnameend}, \bibinfo{author}{Vaishak \surnamestart Belle\surnameend}, \bibinfo{author}{Paolo \surnamestart Felli\surnameend}, \bibinfo{author}{Sheila \surnamestart McIlraith\surnameend}, \bibinfo{author}{Tim \surnamestart Miller\surnameend}, \bibinfo{author}{Adrian~R \surnamestart Pearce\surnameend} \& \bibinfo{author}{Liz \surnamestart Sonenberg\surnameend} (\bibinfo{year}{2015}): \emph{\bibinfo{title}{{Planning Over Multi-Agent Epistemic States: A Classical Planning Approach}}}.
\newblock {\slshape \bibinfo{journal}{Distributed and Multi-Agent Planning (DMAP-15)}}, p.~\bibinfo{pages}{60}, \doi{10.1609/aaai.v29i1.9665}.

\bibitemdeclare{article}{Nau1999}
\bibitem{Nau1999}
\bibinfo{author}{Dana \surnamestart Nau\surnameend}, \bibinfo{author}{Yue \surnamestart Cao\surnameend}, \bibinfo{author}{Amnon \surnamestart Lotem\surnameend} \& \bibinfo{author}{Hector \surnamestart Muftoz-Avila\surnameend} (\bibinfo{year}{1999}): \emph{\bibinfo{title}{{SHOP: Simple hierarchical ordered planner}}}.
\newblock {\slshape \bibinfo{journal}{IJCAI International Joint Conference on Artificial Intelligence}} \bibinfo{volume}{2}, pp. \bibinfo{pages}{968--973}, \doi{10.5555/1624312.1624357}.

\bibitemdeclare{inproceedings}{rao1996agentspeak}
\bibitem{rao1996agentspeak}
\bibinfo{author}{Anand~S \surnamestart Rao\surnameend} (\bibinfo{year}{1996}): \emph{\bibinfo{title}{AgentSpeak (L): BDI agents speak out in a logical computable language}}.
\newblock In: {\slshape \bibinfo{booktitle}{European workshop on modelling autonomous agents in a multi-agent world}}, \bibinfo{organization}{Springer}, pp. \bibinfo{pages}{42--55}, \doi{10.1007/bfb0031845}.

\bibitemdeclare{book}{russell2016artificial}
\bibitem{russell2016artificial}
\bibinfo{author}{Stuart~J \surnamestart Russell\surnameend} \& \bibinfo{author}{Peter \surnamestart Norvig\surnameend} (\bibinfo{year}{2016}): \emph{\bibinfo{title}{Artificial intelligence: a modern approach}}.
\newblock \bibinfo{publisher}{Pearson}.

\bibitemdeclare{article}{Sardina2006}
\bibitem{Sardina2006}
\bibinfo{author}{Sebastian \surnamestart Sardina\surnameend}, \bibinfo{author}{Lavindra \surnamestart {De Silva}\surnameend} \& \bibinfo{author}{Lin \surnamestart Padgham\surnameend} (\bibinfo{year}{2006}): \emph{\bibinfo{title}{{Hierarchical planning in BDI agent programming languages: A formal approach}}}.
\newblock {\slshape \bibinfo{journal}{Proceedings of the International Conference on Autonomous Agents}} \bibinfo{volume}{2006}, pp. \bibinfo{pages}{1001--1008}, \doi{10.1145/1160633.1160813}.

\bibitemdeclare{article}{Schut2002}
\bibitem{Schut2002}
\bibinfo{author}{Martijn \surnamestart Schut\surnameend}, \bibinfo{author}{Michael \surnamestart Wooldridge\surnameend} \& \bibinfo{author}{Simon \surnamestart Parsons\surnameend} (\bibinfo{year}{2002}): \emph{\bibinfo{title}{{On partially observable MDPs and BDI models}}}.
\newblock {\slshape \bibinfo{journal}{Lecture Notes in Computer Science (including subseries Lecture Notes in Artificial Intelligence and Lecture Notes in Bioinformatics)}} \bibinfo{volume}{2403}, pp. \bibinfo{pages}{243--259}, \doi{10.1007/3-540-45634-1_15}.

\bibitemdeclare{article}{Simari2006}
\bibitem{Simari2006}
\bibinfo{author}{Gerardo~I. \surnamestart Simari\surnameend} \& \bibinfo{author}{Simon \surnamestart Parsons\surnameend} (\bibinfo{year}{2006}): \emph{\bibinfo{title}{{On the relationship between MDPs and the BDI architecture}}}.
\newblock {\slshape \bibinfo{journal}{Proceedings of the International Conference on Autonomous Agents}} \bibinfo{volume}{2006}, pp. \bibinfo{pages}{1041--1048}, \doi{10.1145/1160633.1160818}.

\bibitemdeclare{article}{Torreno2014}
\bibitem{Torreno2014}
\bibinfo{author}{Alejandro \surnamestart Torre{\~{n}}o\surnameend}, \bibinfo{author}{Eva \surnamestart Onaindia\surnameend} \& \bibinfo{author}{{\'{O}}scar \surnamestart Sapena\surnameend} (\bibinfo{year}{2014}): \emph{\bibinfo{title}{{FMAP: Distributed cooperative multi-agent planning}}}.
\newblock {\slshape \bibinfo{journal}{Applied Intelligence}} \bibinfo{volume}{41}(\bibinfo{number}{2}), pp. \bibinfo{pages}{606--626}, \doi{10.1007/s10489-014-0540-2}.
\newblock \eprint{1501.07250}.

\bibitemdeclare{article}{VanDrimmelen2023PAL}
\bibitem{VanDrimmelen2023PAL}
\bibinfo{author}{Hans \surnamestart Van~Ditmarsch\surnameend} (\bibinfo{year}{2023}): \emph{\bibinfo{title}{To be announced}}.
\newblock {\slshape \bibinfo{journal}{Information and Computation}} \bibinfo{volume}{292}, p. \bibinfo{pages}{105026}, \doi{10.1016/j.ic.2023.105026}.

\bibitemdeclare{article}{VanDrimmelen2003}
\bibitem{VanDrimmelen2003}
\bibinfo{author}{Govert \surnamestart {Van Drimmelen}\surnameend} (\bibinfo{year}{2003}): \emph{\bibinfo{title}{{Satisfiability in alternating-time temporal logic}}}.
\newblock {\slshape \bibinfo{journal}{Proceedings - Symposium on Logic in Computer Science}}, pp. \bibinfo{pages}{208--217}, \doi{10.1109/lics.2003.1210060}.

\bibitemdeclare{book}{van1995python}
\bibitem{van1995python}
\bibinfo{author}{Guido \surnamestart Van~Rossum\surnameend} \& \bibinfo{author}{Fred~L \surnamestart Drake~Jr\surnameend} (\bibinfo{year}{1995}): \emph{\bibinfo{title}{Python tutorial}}.
\newblock \bibinfo{publisher}{Centrum voor Wiskunde en Informatica Amsterdam, The Netherlands}.

\bibitemdeclare{article}{Vester2013}
\bibitem{Vester2013}
\bibinfo{author}{Steen \surnamestart Vester\surnameend} (\bibinfo{year}{2013}): \emph{\bibinfo{title}{{Alternating-time temporal logic with finite-memory strategies}}}.
\newblock {\slshape \bibinfo{journal}{Electronic Proceedings in Theoretical Computer Science}} \bibinfo{volume}{119}, pp. \bibinfo{pages}{194--207}, \doi{10.4204/eptcs.119.17}.

\bibitemdeclare{article}{Vezina2023}
\bibitem{Vezina2023}
\bibinfo{author}{Michael \surnamestart Vezina\surnameend}, \bibinfo{author}{Fran{\c{c}}ois \surnamestart Schwarzentruber\surnameend}, \bibinfo{author}{Babak \surnamestart Esfandiari\surnameend} \& \bibinfo{author}{Sandra \surnamestart Morley\surnameend} (\bibinfo{year}{2023}): \emph{\bibinfo{title}{{Reasoning About Uncertainty in AgentSpeak Using Dynamic Epistemic Logic}}}.
\newblock {\slshape \bibinfo{journal}{Proceedings of the International Joint Conference on Autonomous Agents and Multiagent Systems, AAMAS}} \bibinfo{volume}{2023-May}, pp. \bibinfo{pages}{2394--2396}, \doi{10.5555/3545946.3598945}.

\bibitemdeclare{article}{Walczak2007}
\bibitem{Walczak2007}
\bibinfo{author}{Andrzej \surnamestart Walczak\surnameend}, \bibinfo{author}{Lars \surnamestart Braubach\surnameend}, \bibinfo{author}{Alexander \surnamestart Pokahr\surnameend} \& \bibinfo{author}{Winfried \surnamestart Lamersdorf\surnameend} (\bibinfo{year}{2007}): \emph{\bibinfo{title}{{Augmenting BDI agents with deliberative planning techniques}}}.
\newblock {\slshape \bibinfo{journal}{Lecture Notes in Computer Science (including subseries Lecture Notes in Artificial Intelligence and Lecture Notes in Bioinformatics)}} \bibinfo{volume}{4411 LNAI}, pp. \bibinfo{pages}{113--127}, \doi{10.1007/978-3-540-71956-4_7}.

\end{thebibliography}

% \vfill
% \noindent\mynote{1}{\textbf{\underline{Targeted venue}}: GandALF2024
% \begin{itemize}
%     \item \textit{Abstract Deadline}: April 16, 2024 (AoE)
%     \item \textit{Paper Deadline}: April 19, 2024 (AoE)
%     \item \textit{Page Limit}: 14 pages + references and appendices (EPTCS)
%     \item \url{https://scool24.github.io/GandALF/}
% \end{itemize}
% }

\end{document}